\documentclass[aps,prb,preprint,preprintnumbers,amsmath,amssymb,superscriptaddress]{revtex4-2}

\usepackage{xr}
\usepackage[utf8]{inputenc}
\usepackage{amsmath}
\usepackage{amsmath,amssymb}
\usepackage{siunitx}
\usepackage[dvipsnames]{xcolor}
\usepackage{color,soul}
\setstcolor{red}
\sethlcolor{yellow}

\newenvironment{eqs}%
{\begin{equation} \begin{aligned}}%
{\end{aligned} \end{equation} }
\newcommand{\beal}{\begin{eqs}}
\newcommand{\eal}{\end{eqs}}

\usepackage{graphicx}
\usepackage{dcolumn}
\usepackage{bm}
\usepackage{booktabs}
\usepackage{amssymb}
\usepackage{amsmath}
\usepackage{url}

\begin{document}
\preprint{APS/123-QED}

\title{ Effective stick-slip parameter for structurally lubric 2D interface friction}

\author{Jin Wang}
    \affiliation{International School for Advanced Studies (SISSA), Via Bonomea 265, 34136 Trieste, Italy}
    \affiliation{International Centre for Theoretical Physics (ICTP), Strada Costiera 11,34151 Trieste,Italy}
 
\author{Andrea Vanossi}
    \affiliation{CNR-IOM, Consiglio Nazionale delle Ricerche - Istituto Officina dei Materiali, c/o SISSA, Via Bonomea 265, 34136 Trieste, Italy}
    \affiliation{International School for Advanced Studies (SISSA), Via Bonomea 265, 34136 Trieste, Italy}
\author{Erio Tosatti}
  \email{tosatti@sissa.it}
    \affiliation{International School for Advanced Studies (SISSA), Via Bonomea 265, 34136 Trieste, Italy}
    \affiliation{International Centre for Theoretical Physics (ICTP), Strada Costiera 11,34151 Trieste,Italy}
    \affiliation{CNR-IOM, Consiglio Nazionale delle Ricerche - Istituto Officina dei Materiali, c/o SISSA, Via Bonomea 265, 34136 Trieste, Italy}

\begin{abstract}
The wear-free sliding of layers or flakes of graphene-like 2D materials, important in many experimental systems,  may occur either smoothly or through stick-slip, depending on driving conditions, corrugation, twist angles, as well as edges and defects.
No single parameter has been so far identified to discriminate a priori  between the two sliding regimes. Such a parameter, $\eta$, does exist in the ideal (Prandtl-Tomlinson) problem of a point particle sliding across a 1D periodic lattice potential.
In that case $\eta >1$ implies mechanical instability, generally leading to stick-slip, with $\eta = \frac{2\pi^2 U_0}{K_\mathrm{p} a^2}$, where $U_0$ is the potential magnitude, $a$ the lattice spacing, and $K_\mathrm{p}$ the pulling spring constant.
Here we show, supported by a repertoire of graphene flake/graphene sliding simulations, that a similar stick-slip predictor $\eta_\mathrm{eff}$ can be defined with the same form but suitably defined $U_\mathrm{eff}$, $a_\mathrm{eff}$ and $K_\mathrm{eff}$.
Remarkably, simulations show that $a_\mathrm{eff} = a$ of the substrate remains an excellent approximation, while $K_\mathrm{eff}$ is an effective stiffness parameter, combining equipment and internal elasticity. Only the effective energy barrier $U_\mathrm{eff}$ needs to be estimated in order to predict whether stick-slip sliding of a 2D island or extended layer is expected or not.
In a misaligned defect-free circular graphene sliding island of contact area $A$, we show that $U_\mathrm{eff}$, whose magnitude for a micrometer size diameter is of order 1~eV, scales as $A^{1/4}$, thus increasing very gently with size.
The PT-like parameter $\eta_\mathrm{eff}$ is therefore proposed as a valuable tool in 2D layer sliding.

\end{abstract}

\date{\today}

\maketitle

\section{Introduction}

The contact interface between graphene or graphene-like 2D material layers and flakes or islands has acquired great importance in the last decade \cite{Geim.nature.2013,Bhimanapati.acsnano.2015,Neto.science.2016,Schulman.CSR.2018,Glavin.advmater.2020}. Owing to the great strength of both slider and substrate, an applied planar force can cause this interface to slide without damage or wear \cite{Peng.pnas.2020, Hod.nature.2018}.
Both experiments and simulations have explored the frictional aspects of the sliding process, as reviewed in \cite{Wang.rmp.2024}. In particular, when a 2D island or layer is forced, through a tip or a spring, to slide on a substrate, different frictional behaviours are in principle possible, depending basically on the nature of total free energy  $E(\bm{r})$, generally referred to as $E(x)$, as a function of the relative coordinate $x$ of the two centers of mass.

The first possibility, usually known as structural superlubricity, is academic and strictly applies only to the ideal case where both layers are of infinite size, defect free, incommensurate (and Aubry unpinned \cite{Aubry.jpc.1983,Vanossi.rmp.2013}), is $E(x) = \mathrm{const.}$.
Because there is no energy barrier, superlubric sliding occurs for an arbitrarily weak applied force, with tiny frictional dissipation -- mostly due to moir\'e out-of-plane motions \cite{Wang.jmps.2023,Song.natmater.2018,Mandelli.prl.2019} -- proportional to velocity. 
The second possibility, more realistic even if uncommon in practice, is realized when flake edges, defects, or weak commensurability cause $E(x)$ to depend on $x$, but the effective barrier $U_\mathrm{eff}= \max \{\Delta E(x)\} = \max \{E(x)\} - \min \{E(x)\}$ is weak relative to a hard pulling spring whose stiffness $K_\mathrm{p}$ is large. 
In this case too one may still have smooth sliding, with the average value of the washboard oscillating frictional force still growing linearly with velocity \cite{Wang.rmp.2024}.
The third and commonest case occurs when the free energy barrier $U_\mathrm{eff}$ is strong, and/or the pulling spring $K_\mathrm{p}$ is soft. In this case the sliding motion can only occur through a succession of mechanical instabitities and, as in the one-dimensional Prandtl-Tomlinson (PT) model \cite{Persson.2000} stick-slip will ensue.
The average stick-slip friction force in this case remains large even at low velocities, and its growth with velocity becomes much weaker, typically logarithmic rather than linear \cite{Wang.rmp.2024,Peng.pnas.2020,Song.natmater.2018,Wang.epl.2019,Liu.acsami.2020,Li.nanoscale.2020}.\\

With the last two realistic situations of nonzero barrier in mind, we are concerned here with understanding and possibly predicting the occurrence of either smooth sliding or stick-slip ahead of experiments and without recourse to simulations. A concise parameter that could discriminate between two sliding states is clearly desirable.
In the paradigmatic 1D PT model, where the total potential energy is
$E(x,t) = \frac{U_0}{2} \cos(\frac{2\pi x}{a}) + \frac{K_\mathrm{p}}{2} (x -x_\mathrm{spring})^2$
there is precisely such a parameter,
\begin{equation}
    \eta=\frac{2\pi^2 U_0}{K_\mathrm{p} a^2}
\end{equation}
where $K_\mathrm{p}$ is the pulling spring stiffness, $a$ the periodic potential spacing, and the energy barrier $U_0$ is the potential magnitude \cite{Socoliuc.prl.2004, Vanossi.rmp.2013}.
In this model, {mechanical stability},
$\partial^2 E/\partial x^2 >0$,  occurs for $\eta < 1$, a situation verified when the barrier is weak and the spring is stiff, the mechanical evolution is stable and the sliding motion is smooth. For $\eta > 1$, the evolution encounters mechanical instability and the sliding develops discontinuities, which give rise to stick-slip.
Simple as it is this model and $\eta$ parameter describes well the transition between smooth sliding and stick-slip of tip-based frictional systems, as also verified by a variety of experiments \cite{Socoliuc.prl.2004, Medyanik.prl.2006,Szlufarska.jpd.2008,Roth.tribolett.2010,Buzio.carbon.2021,Buzio.acsami.2023}.

\begin{figure*}[ht!]
\centering
\includegraphics[width=0.8\linewidth]{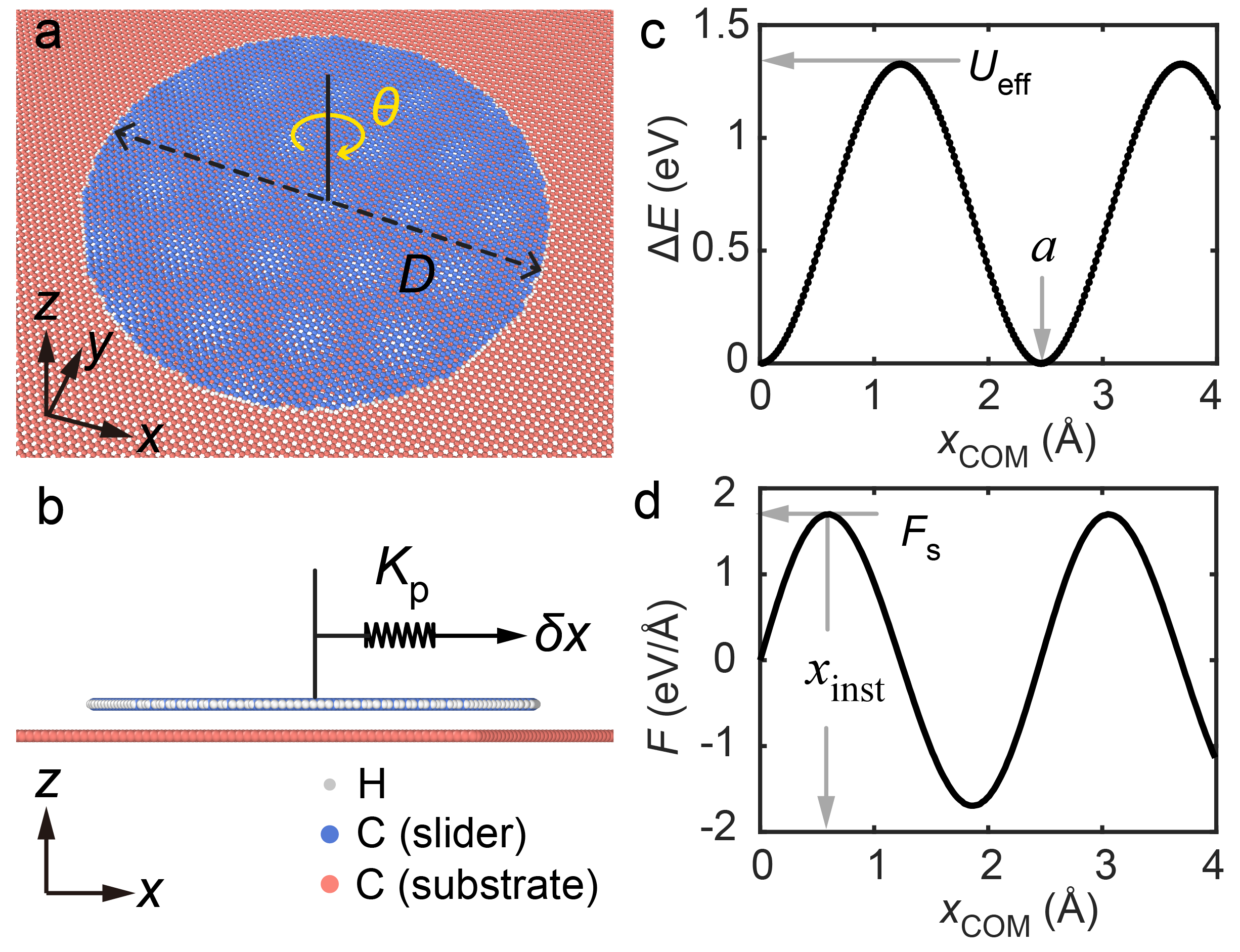}
\caption{Models and protocols.
(a) Schematic sketch of the simulation model. The slider (blue) has a diameter $D$ and rotated by $\theta$ degree with respect to the rigid substrate (pink). The $x$-axis is paralleled to the zigzag direction of the substrate.
(b) Side view of the model and simulation protocol.
(c) The sliding energy landscape and (d) the lateral force trace.
The slider used here has a diameter $D=14$~nm and twist angle $\theta=5^{\circ}$.
The lattice constant of the graphene substrate is $a=\sqrt{3} l_\mathrm{CC}$, where $l_\mathrm{CC}=1.42~\mathrm{\AA}$ is the equilibrium C-C bond length.}
\label{fig:model}
\end{figure*}

We are interested here in extending this kind of parameter to 2D structurally lubric (SL) systems, such as mesoscale size islands sliding on crystalline substrates in incommensurate contact \cite{Liu.prl.2012,Li.advmater.2017,Song.natmater.2018,Liao.natmater.2022}. A sliding flake or island is in principle a much more complex system, encompassing a larger number of degrees of freedom as opposed to just one as in the PT model (Fig.~\ref{fig:model}b). The strength of bonds in a 2D material however enslaves all atomic coordinates of the island, at least during adiabatic, quasi-static sliding, to just macroscopic coordinates, namely the center-of-mass (COM) coordinate $\vec{R}$, plus the island-substrate ``twist" angle $\theta$. In  many practical cases, moreover, the island is forced to slide by drivers that cannot rotate.   With this situation in mind, it seems natural to try to identify an $\eta$  parameter also for extended 2D SL contacts.

Unsurprisingly, this kind of extension requires great caution, with many issues and complications with respect to the PT model. Even without rotations, the potential field of the 2D surface-to-surface contact is generally vastly different from sinusoidal. It will depend on small internal elastic distortions, both in-plane and out-of-plane, that accompany the COM motion.
Other features, including island size, twist angle $\theta$, sliding direction $\Phi$, slider shape, etc., will act to deform the potential field. 
With these caveats in mind, we may still tentatively submit to test a trial $\eta_\mathrm{eff}$ with the same PT form but where all relevant parameters $U, K, a$ can take effective magnitudes that differ case by case. The eventual quality of this trial remains to be determined and judged by discovering what values these constituent parameters take in practice -- a task we propose to pursue here by realistic simulations. We thus propose to try
\begin{equation}
    \eta_\mathrm{eff} = \frac{2\pi^2 U_\mathrm{eff}}{K_\mathrm{eff} a_\mathrm{eff}^2}
\end{equation}
As said above the effective free energy barrier (inclusive of temperature effects if present) is $U_\mathrm{eff} = \max \{\Delta E(x)\}$ along the chosen sliding direction. More delicate and crucial is the definition of effective substrate periodicity $a_\mathrm{eff}$. We propose using 
\begin{equation}
 a_\mathrm{eff}=4 x_\mathrm{inst}  
\end{equation}
where $x_\mathrm{inst}$ is the COM coordinate where mechanical instability will occur upon sliding. When the island is displaced from its equilibrium at $x=0$ to $x_\mathrm{inst}$, mechanical instability occurs when the second derivative of $E(x)$ changes sign for at least one value of $x_\mathrm{spring}$.
That is also the point of maximum lateral force, $dE/dx|_{x=x_\mathrm{inst}} = \max \{dE/dx \}$. 
Finally, the effective stiffness $K_\mathrm{eff}$ is generally affected by the internal elastic stiffness $K_\mathrm{slider}$ of the slider, typically in the 
spring chain form
\begin{equation}
    K_\mathrm{eff}^{-1}=K_\mathrm{p}^{-1}+K_\mathrm{slider}^{-1}.
\label{eq3}
\end{equation}
The key problem to be answered at this stage is therefore, how predictable or unpredictable these three effective parameters might be in practice.
That is, how large is their difference from those that could be just guessed, e.g., by treating the whole island as a point slider.\\

In the rest of this work, by using mainly twisted graphene islands as our demonstration workhorse, we employ molecular dynamics simulations to study systematically the sliding energy landscape for cases with different contact areas, twist angles, sliding directions, edge shapes, lattice mismatches, pinning defects.
Devoid of conceptual pretense, the work aims at providing a practical tool that could predict the quasi-static sliding behaviour of structurally lubric sliders.
Parameters $U_\mathrm{eff}$ and $a_\mathrm{eff} $ are estimated from case to case to discover if/how they may reasonably represent the sliding of a generic SL island.
The influence of contact elasticity, absent in the PT model (and still negligible as we shall see for most nanoscale islands), should not in general be forgotten. Elasticity must play an important a role in contacts exceeding the micron size, in which case the effective stiffness will diminish relative to the bare external pulling stiffness, as suggested by Eq.~(4) above. 
Further, we will discuss the impact of common defects existing in real systems, beyond the island perimetral edges that provide the omnipresent sliding energy barrier. In conclusion, we will show how to make use of $ \eta_\mathrm{eff}$ in order to seek experimental conditions that will minimize or maximize friction .

\section{Simulations: Model and Methods}

In simulations we focus on statics of the slider-substrate interface, as appropriate to ascertain the nature of static friction (smooth versus mechanically unstable). Kinetic friction simulations of similar models can be found e.g., in ref.~\cite{Wang.rmp.2024}.
Our main MD simulation model consists of a rigid graphene substrate (also a rigid Au(111) substrate in Section VI) with a finite-sized graphene slider, 
initially rigid (in Section III to VI), then fully flexible in subsequent Sections VII to IX, portrayed in Fig.~\ref{fig:model}(a, b).
We focus on a circular slider shape with diameter $D$. The effects of different shapes will be discussed in Section V. The edge of the slider is passivated by H-atoms. The slider is generally rotated by a twist angle $\theta$ with respect to the substrate.
In order to keep this exploration at the simplest level, temperature was set throughout at $T=$0. Because we wish to address large sliders, all that can change at $T >$ 0 is a possible thermolubric reduction of edge- or defect-related free energy barriers , purely quantitative and of decreasing relevance as the island size increases.

All simulations are performed with the LAMMPS code \cite{Plimpton.jcp.1995,Thompson.compphyscomm.2022}. The interlayer and intralayer interaction are described by REBO and  by registry-dependent ILP force fields respectively \cite{Brenner.jpcm.2002,Ouyang.nanolett.2018, Ouyang.jctc.2021}.
Without attempting to mimic the actual experimental forcing, the center-of-mass of the slider is dragged by a moving spring of stiffness $K_\mathrm{p}$.
For simulations with rigid flakes (Section III to VI), the total potential energy $E(x)$ is scanned by rigidly displaced the slider. For convenience, here we focus on $\Delta E(x) = E(x) -\min\{E\}$ (shown in Fig.~\ref{fig:model}c).
The lateral (driving) force is calculated by $F=dE/dx$ (Fig.~\ref{fig:model}d). Effective sliding free energy barrier and static friction are defined by
$U_\mathrm{eff} = \max \{\Delta E(x)\}$ and
$F_\mathrm{s} = \max \{F(x)\}$ respectively.
For simulations with flexible islands (Section VII to IX), a quasi-static sliding protocol is adopted (Fig.~\ref{fig:model}b). Starting from an energy minimum, the slider is displaced by a pulling spring with spring constant
$K_\mathrm{p}$, a parameter controlled by the external driving system \cite{Szlufarska.jpd.2008, Dong.jvst.2013}. In AFM-based experiments, its magnitude is typically within the range of $1 \sim 100$~N/m \cite{Dienwiebel.prl.2004,Liu.rsi.2007,Song.natmater.2018,Liao.natmater.2022}.
During the quasi-static sliding, one end of the pulling spring is tethered to the COM of the slider, the other end is displaced by $\delta x=0.02~\mathrm{\AA}$ in each step,
followed by a full structural optimization with the CG+FIRE algorithm. The energy and force tolerance used in optimizations are $10^{-15}$ and $10^{-4}$~$\rm{eV/\AA}$ respectively. 
To restrict the global rotation of the slider, planar springs perpendicular to the sliding direction (with spring constant $k_i=1$~N/m) are tethered to each slider atom -- a virtual constraint to counteract the global torque \cite{Wang.rmp.2024}.

\section{Slider size and twist angle}

\begin{figure*}[ht!]
\centering
\includegraphics[width=1\linewidth]{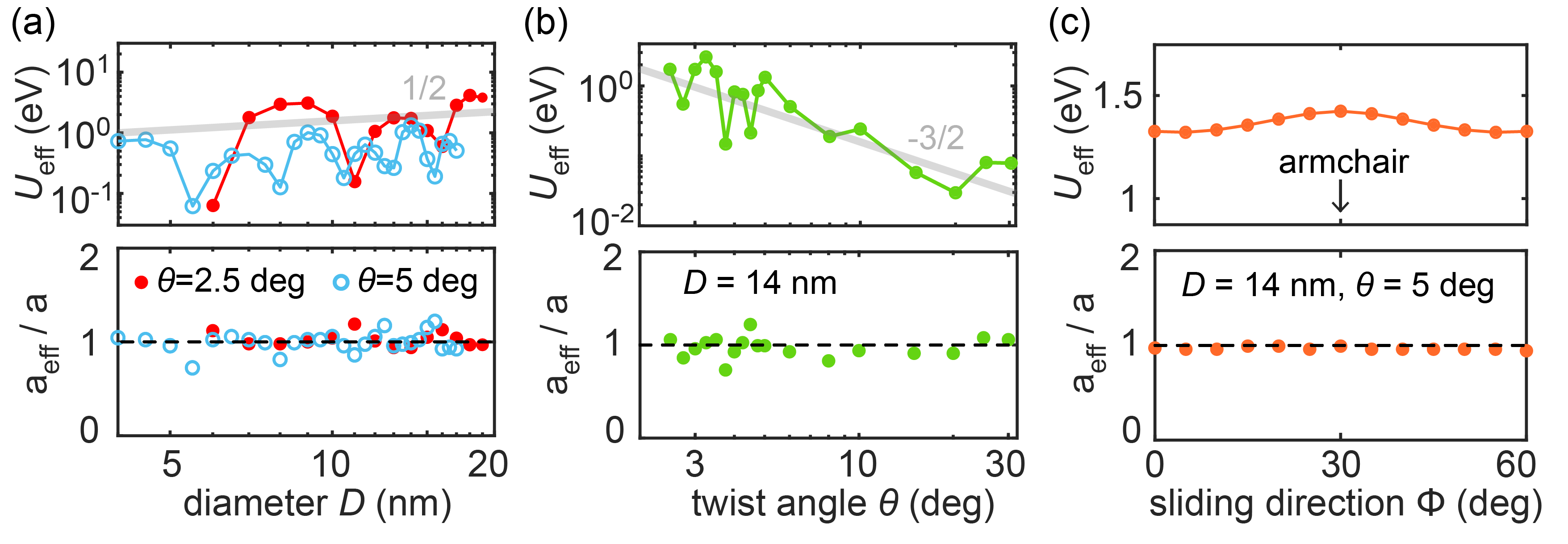}
\caption{Effective energy barrier $U_\mathrm{eff}$ (upper panels) and periodicity $a_\mathrm{eff}$ (lower panels) as a function of (a) diameter $D$, (b) twist angle $\theta$, and (c) sliding direction $\Phi$. The power function fits and the scaling exponents are shown in (a) and (b).
}
\label{fig:basic}
\end{figure*}

With the simulation protocol introduced above, we are 
set to discuss the influence to $U_\mathrm{eff}$ and $a_\mathrm{eff}$ from various factors. We begin in this section with the size and twist angles.
Simulations results for the size and twist angle dependence of $U_\mathrm{eff}$ and $a_\mathrm{eff}$ are shown in Fig.~\ref{fig:basic} (a-b).\\

{\it Size dependence.} The values of $a_\mathrm{eff}$ are found to be uniformly close to $a$, independent of size, and as we shall see later, approximately independent of driving details such as side or central pulling.
The energy barrier of the island $U_\mathrm{eff}$, due to the edges which even in the absence of other defects break full translational symmetry, logically increases with size, as does the perimeter. Physically the barrier is due to the uncompensated moir\'e nodes entering/exiting the edge.
Its size scaling for a SL system is $U_\mathrm{eff} \sim U_i (D/a_\mathrm{sli})^{\gamma}$, where $U_i$ represents the per-atom sliding energy barrier, $a_\mathrm{sli}$ is the lattice constant of the slider and $\gamma$ is a scaling exponent.
For a {defect-free} graphene/graphene interface, the basic parameter  $U_i$ determining the edge-induced energy barrier is estimated with the present force field to be about $0.1$~eV.
By fitting the upper envelope of simulation results (Fig.~\ref{fig:basic}a), we get $\gamma \sim 1/2$, i.e., the barrier is approximately proportional to the slider perimeter's square root.
This scaling, $U_\mathrm{eff} \propto A^{1/4} \propto D^{1/2}$ ($A$ is the contact area), agrees with previous studies of circular islands \cite{Koren.prb.2016,Sharp.prb.2016,Wang.nanolett.2019, Wang.rmp.2024,Yan.arxiv.2023}.
Its meaning is that among all perimetral atoms, only the front and rear ones dominate the friction, a fact also well established for nanoribbons \cite{Kawai.science.2016, Gigli.2dmater.2017, Wang.rmp.2024}. That implies as we will show in Sect V that $\gamma$ is shape dependent, and can generally rise from 0 to 1.

{\it Twist dependence.}
The values of $a_\mathrm{eff}$ remain close to $a$, for all twist angles $\theta$ we studied here, from $2^{\circ}$ to $30^{\circ}$.
The effective barrier $U_\mathrm{eff}$ has a more interesting dependence on twist angle. The largest value $U_i \sim 0.1$~eV is obtained for systems with small twist angles.
The barrier drops as $\theta$ increases (Fig.~\ref{fig:basic}b), scaling as $U_\mathrm{eff} \propto \theta^{\gamma-2}$, a decrease due to the decreasing contribution from the moir\'e edge \cite{Koren.prb.2016}.
For a system with $\theta=30^{\circ}$ -- {closest to the} ideal ``superlubric'' state, its $U_i$ value is even smaller, approximately $10^{-2}$~eV.

\section{Sliding direction}

The next point that distinguishes the real-world SL system from the 1D PT model is the sliding direction -- the potential energy evolves differently as the COM of the slider moves along different directions (characterized by the angle $\Phi$ with $x$-axis).
By symmetry, both $U_\mathrm{eff}$ and $a_\mathrm{eff}$ possess $60^{\circ}$ symmetry with respect to $\Phi$.

Simulation results of a $D=14$~nm and $\theta=5^{\circ}$ model are shown in Fig.~\ref{fig:basic} (c). For sliding directions from $\Phi = 0^{\circ}$ to $60^{\circ}$, both $U_\mathrm{eff}$ and $a_\mathrm{eff}$ do not differ much.
Once again, for all the graphene islands which we considered, with diameter $D$ ranging from 4 to 20 nm and $\theta$ from $2^{\circ}$ to $30^{\circ}$, the effective periodicity $a_\mathrm{eff}$ was found to remain remarkably close to the substrate (also graphene) lattice constant $a$. The overall shape of $E(x)$ being very strongly dependent upon the sliding direction $\Phi$, this result seems quite surprising. It is explained as follows.

Assuming a weak interaction between slider and substrate \cite{deWijn.prb.2012,Sharp.prb.2016,Wang.nanolett.2019,Wang.jmps.2023}, the general $E(\bm{r})$ can be represented by
\begin{equation}
    E(\bm{r})= - \frac{2 U_\mathrm{slider}}{9} \left[ \sum_{i=1}^{3} \cos(\bm{k}_i \cdot \bm{r}) +\frac{3}{2} \right ]
    \label{2d_potential}
\end{equation}
where $U_\mathrm{slider}$ is the barrier of the whole slider and $\bm{k}_i$ is the reciprocal vector of the triangular lattice, with magnitude
$|\bm{k}_i| = 4\pi/\sqrt{3} a$, and $a$ is the lattice constant of the substrate.

Starting from the energy minimum, i.e., $\bm{r}=0$, one can get the slope along an arbitrary direction $\Phi$,
\begin{equation}
    S=\nabla E \cdot \vec{t}
\end{equation}
where $\vec{t}=(\cos \Phi, \sin \Phi)$.
The steepest slope along direction $\Phi$ satisfies
\begin{equation}
    \frac{dS}{dr} = 0
\end{equation}
In polar coordinates where $x = r \cos{\Phi}$ and $y = r \sin{\Phi}$, one gets
\begin{equation}
\begin{aligned}
    \frac{dS}{dr} &= \frac{8\pi^2 U_\mathrm{slider}}{27a^2} 
    \big \{
    \cos[\frac{2\pi r}{3a} (\sqrt{3} \cos \Phi -3\sin \Phi)]
    (\cos\Phi-\sqrt{3}\sin\Phi)^2 \\
    &+ 
    \cos[\frac{2\pi r}{3a} (\sqrt(3) \cos \Phi +3\sin \Phi)]
    (\cos\Phi+\sqrt{3}\sin\Phi)^2 \\
    &+
    4\cos(\frac{4\pi r}{\sqrt{3} a} \cos\Phi) \cos^2\Phi
    \big \}
\end{aligned}
\end{equation}
With a variable substitution:
\begin{equation}
\begin{aligned}
    m &= \cos\Phi-\sqrt{3}\sin\Phi \\
    n &= \cos\Phi+\sqrt{3}\sin\Phi
\end{aligned}
\end{equation}
the above formula simplifies to
\begin{equation}
\begin{aligned}
    \frac{dS}{dr} &= \frac{8\pi^2 U_\mathrm{slider}}{27a^2} 
    \big \{
    \cos(\frac{2\pi m r}{\sqrt{3} a}) m^2 + 
    \cos(\frac{2\pi n r}{\sqrt{3} a}) n^2 +
    \cos[\frac{2\pi r}{\sqrt{3} a} (m+n)] (m+n)^2
    \big \}
\end{aligned}
\end{equation}
Considering the 60-degree symmetry of $\Phi$ and the fact that the largest slope position must be inside the potential well, the cosine terms can be approximated as $\cos{x} \approx 1-x^2/2+x^4/24$.
The above equation further simplifies to
\begin{equation}
\begin{aligned}
    \frac{dS}{dr} &= \frac{8\pi^2 U_\mathrm{slider}}{27a^2} 
    \big \{ [m^2+n^2+(m+n)^2] - \frac{2\pi^2 r^2}{3a^2}[m^4+n^4+(m+n)^4] \\
    &+\frac{2\pi^4 r^4}{27a^4} [m^6+n^6+(m+n)^6] + O[{(\frac{r}{a})}^6]
    \big \}
\end{aligned}
\end{equation}
Noting that
\begin{equation}
\begin{aligned}
    m^2&+n^2+(m+n)^2 = 6  \\
    m^4&+n^4+(m+n)^4 = 18 \\
    m^6&+n^6+(m+n)^6 = 60+6\cos(6\Phi)
\end{aligned}
\end{equation}
Substituting Eqs.~(12) into Eq.~(11), we finally conclude that
\begin{enumerate}
    \item $dS/dr$ is weakly $\Phi$-dependent (when $r<a$);
    \item $dS/dr=0$ occurs at $r\approx a/4$.
\end{enumerate}

In simple words, even though the overall $E(\bm{r})$ is strongly direction dependent, its quadratic growth near $r=0$ is approximately independent of direction, and so is the instability point of maximum slope. That implies that case-by-case corrections to $a_\mathrm{eff}$ are unnecessary, this parameter being well approximated by the bare substrate lattice constant.
Given this weak directional dependence, simulations in the subsequent sections are all along the zigzag ($x$) direction.

The effective barrier does, unlike $a_\mathrm{eff}$, depend upon the sliding direction, although only weakly. The sixfold symmetry is confirmed in $U_\mathrm{eff}(\Phi)$, and the relative difference of $U_\mathrm{eff}$ between zigzag ($\Phi=0$) and armchair ($\Phi=30^\circ$) sliding direction is only $\approx 10\%$, a difference which also agrees with experimental sliding of SL graphite/hBN interfaces \cite{Song.prm.2021}.

\section{Slider Shape}

\begin{figure*}[ht!]
\centering
\includegraphics[width=1\linewidth]{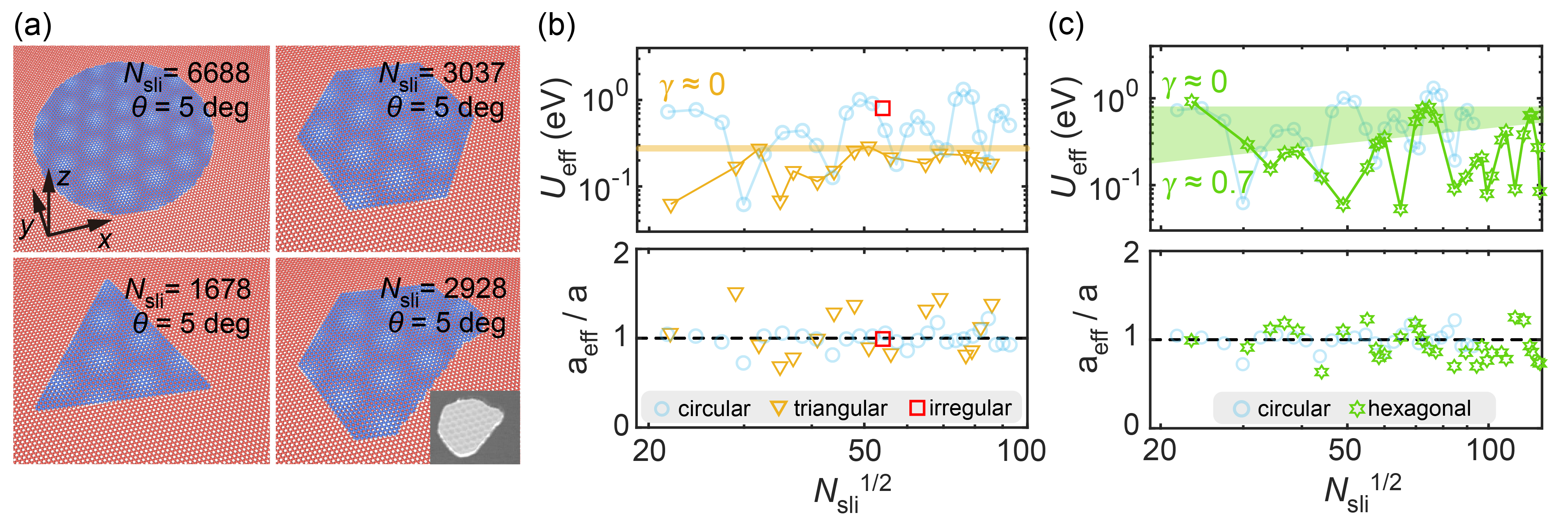}
\caption{Results for different slider shapes.
(a) Simulation model with circular, hexagonal, triangular, and an irregular shape slider. This shape is built based on the shape of flake reported in experiments (inset) \cite{Liao.natmater.2022}. Number of slider atom $N_\mathrm{sli}$ is shown in the figure. Twist angle $\theta= 5^{\circ}$ for all shapes in the simulation.
(b, c) Effective energy barrier $U_\mathrm{eff}$ and periodicity $a_\mathrm{eff}$ as a function of the contact size $N_\mathrm{sli}^{1/2}$.
For clarity, (b) and (c) show the results for triangular (yellow) and irregular-shaped (red) islands and hexagonal islands (green) respectively. Circular shape results (blue) are shown in background to facilitate direct comparison.
The fitting and the corresponding scaling exponents $\gamma$ are shown in the upper panels of (b) and (c). The exponent for the hexagonal shaped case scatters, from 0 to $\approx 0.7$.}
\label{fig:shape}
\end{figure*}

Besides circular sliders \cite{Koren.prb.2016, Li.advmater.2017, Song.prm.2021}, instructive if not particularly realistic, there are many other candidate shapes that may help anticipate the often irregular forms encountered in 2D material-based SL experiments \cite{Dietzel.prl.2013,Ozogul.apl.2017,Liao.natmater.2022}.
In this section, we examine results for triangular, hexagonal, and  mixed-shape sliders -- to explore the variety of cases (Fig.~\ref{fig:shape}a). In view of different shapes, we use the number of carbon atoms in the slider $N_\mathrm{sli}$ to characterize the size of the model. A circular slider with $N_\mathrm{sli}=6688$ has diameter $D=15$~nm.
The edges of triangular and hexagonal sliders used in our simulations are along zigzag directions, a choice based on the fact that zigzag direction has a slightly lower fracture toughness \cite{Qu.prl.2022}. Nonetheless, the orientation of edges in experiments could still scatter. This means that even for the same shape, the orientation of the edges can further affect the result \cite{Wang.nanolett.2019}.

Similar to previous observations, simulation results in Fig.~\ref{fig:shape}(b) show that $a_\mathrm{eff}$ remains close to $a$ for all shapes and size we studied, even for the irregular shape case (marked by red squares).
The effective energy barriers are smaller for triangular and hexagonal systems than for circular shapes, at least for the chosen $\theta=5^{\circ}$ and $\Phi=0$. A larger barrier occurs when more moir\'e nodes simultaneously cross the island edge, and coincident crossings happen to be less abundant in the chosen shapes compared to circular. The barrier growth with size is also sublinear.
Even if data are insufficient to extract an accurate scaling exponent $\gamma$ from $ U_\mathrm{eff} \sim N_\mathrm{sli}^{\gamma/2}$, the data are compatible with $0<\gamma <1$ as anticipated.
We note that certain shapes show a surprising $\gamma <1/2 $, such as the triangular shape of Fig.~\ref{fig:shape}(b), where $\gamma \approx 0 $.
As also seen in some previous simulations \cite{Wang.nanolett.2019,Yan.arxiv.2023}, this surprising lack of growth of the sliding barrier with size is possible when polygonal islands slide along or close to a {\it wedge} direction. This reinforces the concept that a choice of shapes and orientation might be crucial when seeking structurally lubric sliding of large size sliders.

\section{Heterostructures}

\begin{figure*}[ht!]
\centering
\includegraphics[width=0.8\linewidth]{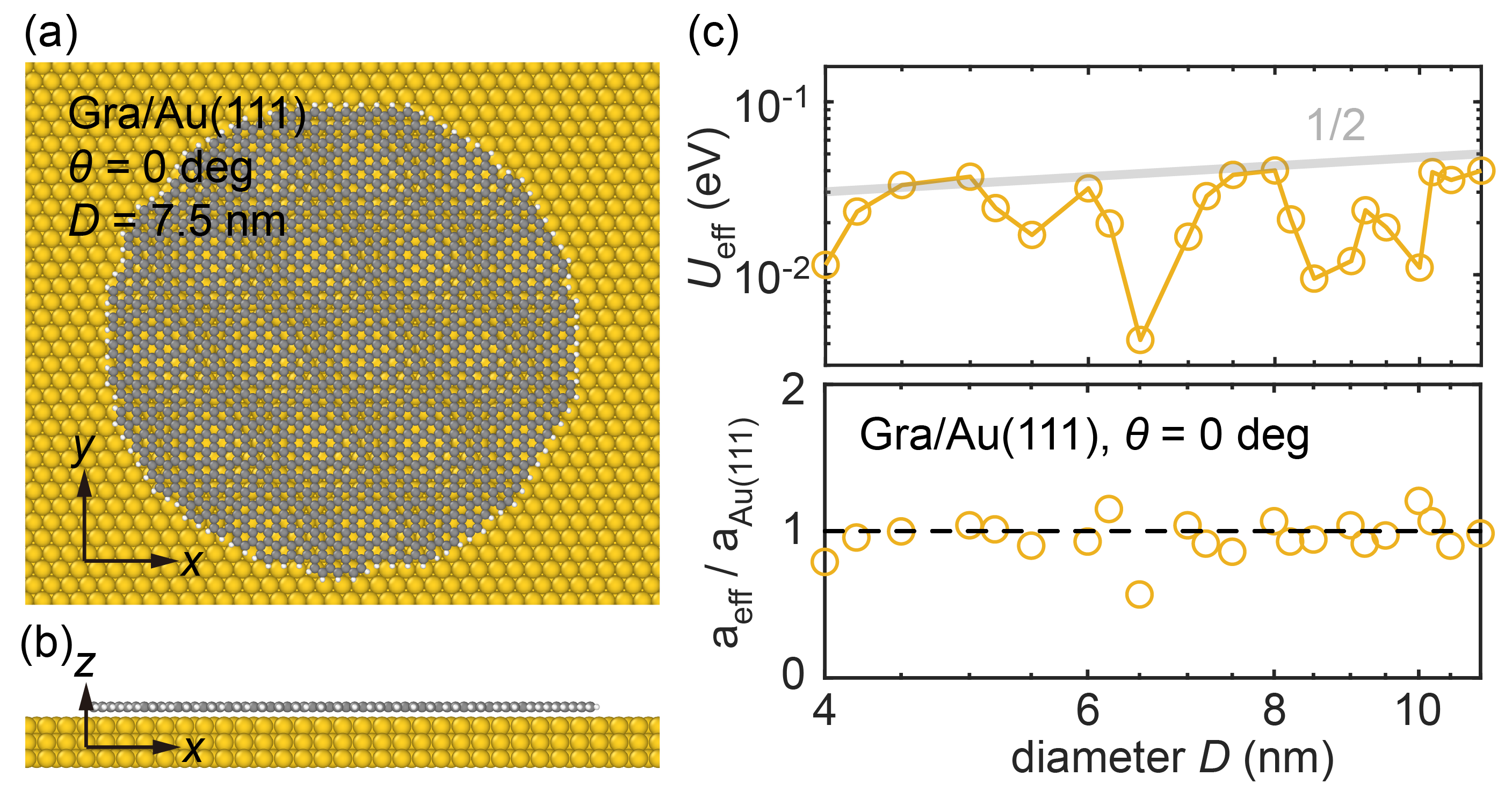}
\caption{Models and results for graphene/Au(111) heterostructure.
(a, b) Top view and side view of the simulation model.
(c) Effective energy barrier $U_\mathrm{eff}$ and periodicity $a_\mathrm{eff}$ as a function of the diameter of the slider. Twist angle $\theta= 0^{\circ}$ (aligned) for all heterostructures used here. The lattice spacing of the Au(111) substrate is $a_\mathrm{Au(111)}\approx2.885$~\AA.
}
\label{fig:hetero}
\end{figure*}

The above results and discussion was focused on graphene homo-structures. Although graphene and its interfaces are still most popular in SL research, hetero-structures are attracting increasing interest.
That is because of their rich electronic properties \cite{Geim.nature.2013,Bhimanapati.acsnano.2015,Neto.science.2016} but also of their robust ``superlubric'' behavior -- the sliding energy barrier remains low under arbitrary twist angles \cite{Leven.jpcl.2013,Dietzel.prl.2013,Gigli.2dmater.2017,Song.natmater.2018}.
In this section, we consider the aligned graphene/Au(111) hetero-structure (Fig.~\ref{fig:hetero}a, b) as a representative hetero-interface. We simulate this system to extract the size dependence of $U_\mathrm{eff}$ and $a_\mathrm{eff}$, giving a direct comparison to the results in Section III.

Simulation results in Fig.~\ref{fig:hetero}(c) show that the main conclusions in previous sections still hold for hetero-structures. Specifically, $U_\mathrm{eff}$ scales sublinearly with diameter ($\gamma=1/2$), and $a_\mathrm{eff}$ is very close to the lattice constant of the substrate  $a_\mathrm{Au(111)}=a_\mathrm{Au}/\sqrt{2}\approx2.885$~\AA, where $a_\mathrm{Au}=4.08~\mathrm{\AA}$ is the lattice constant of gold.
In addition, the magnitude of the sliding energy barrier of graphene/Au(111) hetero-junction is tiny -- comparable to the graphene homo-structure with large twists (Fig.~\ref{fig:basic}), which shows that the system has exceptionally good superlubric properties \cite{Kawai.science.2016,Li.carbon.2020,Ouyang.jctc.2021}.

\section{Elasticity}

\begin{figure*}[ht!]
\centering
\includegraphics[width=0.8\linewidth]{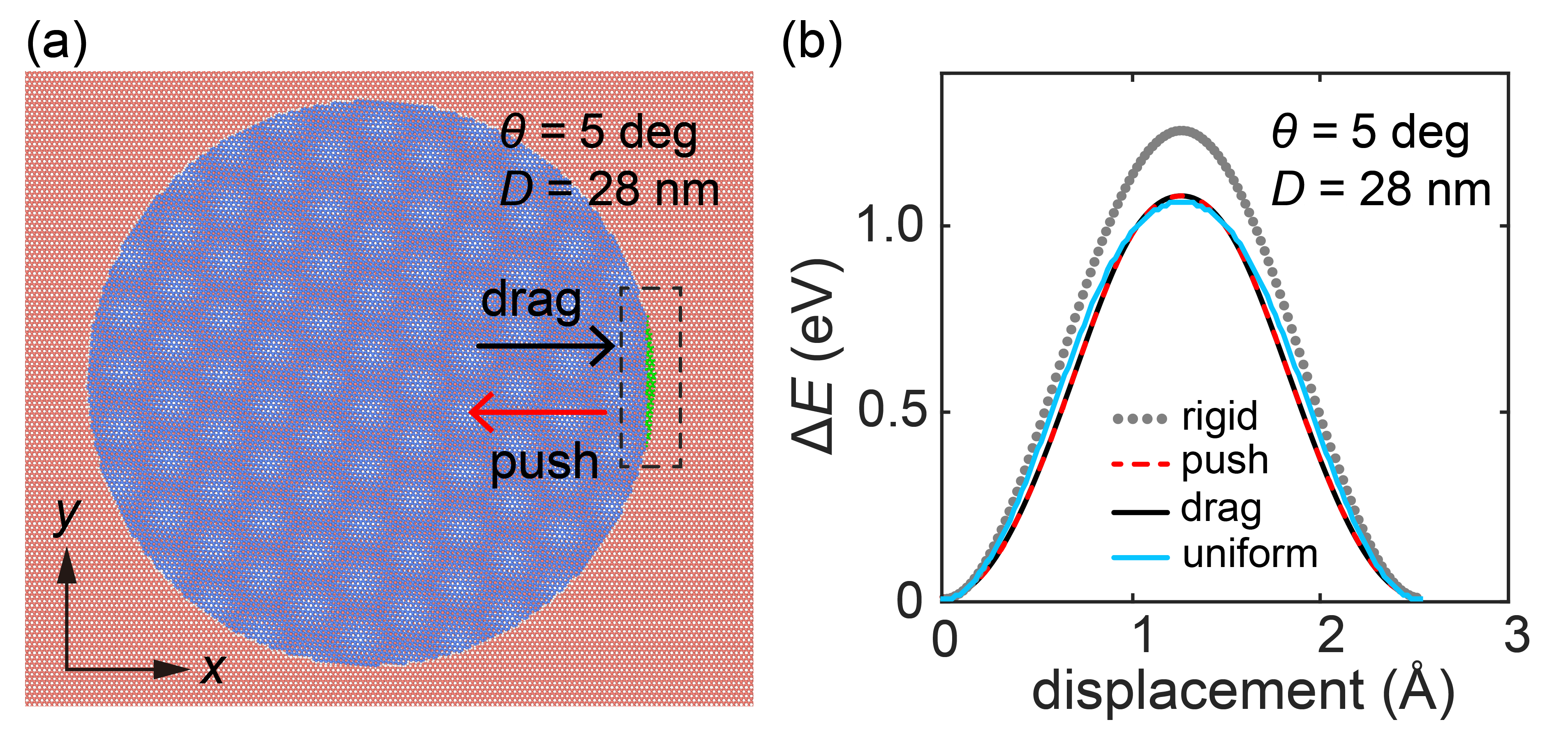}
\caption{Models and results for different driving methods. (a) Schematic sketch of the simulation model. In edge-drag and edge-push simulation, the slider moves by dragging/pushing the rightmost edge (highlighted green). 
(b) Potential energy evolution as a function of displacement.
Parameters used in simulations are $D=28$~nm ($N_\mathrm{sli} = 23437$) and $\theta=5^{\circ}$.
}
\label{fig:flexibility}
\end{figure*}

The simulations in the previous four Sections are based on rigid {island} sliders. The rationale for this choice is that 2D materials such as graphene are very stiff. However, as size increases, or when the driving method changes, the influence of elasticity may no longer be ignored.
In this section, the simulated graphene island sliding on a rigid graphene substrate is flexible with $D=28$~nm and $\theta=5^{\circ}$.
To compare with the rigid island case, three simulation protocols are introduced, namely, uniform drag, edge-drag and edge-push, corresponding to three typical driving methods in experiments \cite{Dietzel.prl.2009,Kawai.science.2016,Song.natmater.2018,Liao.natmater.2022,Yang.acsami.2023}.
For the edge drag and push cases, the pulling spring is tethered to the narrow edge region (green color in Fig.~\ref{fig:flexibility}a) instead of the COM as in the uniform case.

Results are shown in Fig.~\ref{fig:flexibility}(b). The flexibility of the island, both in-plane and out-of-plane, causes the potential energy $E(x)$ to decrease compared to the rigid case at all positions, and to further deviate from sinusoidal.
Due to that, the relaxed $a_\mathrm{eff}$ grows slightly larger than $a$. For both edge-driven cases, we find $a_\mathrm{eff}=2.72~\mathrm{\AA}$, compared to $a=2.46~\mathrm{\AA}$ of the rigid case.
This increase is connected with the  entry and exit of the  moir\'e pattern AA nodes --  higher energy regions \cite{Koren.prb.2016,Cao.prx.2022,Wang.rmp.2024} -- at the edge of the slider.
For a rigid slider, the AA node is forced to enter/exit smoothly from the edge; while for a flexible slider, deformability causes the entry and exit of local AA to 
delay the overall COM movement of the slider -- the AA node remains pinned at the edge for a while.
This pinning cannot last long, especially for 2D materials with very high in-plane stiffness -- and when the next moiré is about to approach the edge, the previous pinned moiré is forced to leave. As a result, although elasticity causes $a_\mathrm{eff} > a$, that increase is always modest, well below its sinusoidal upper limit $2a$.
The potential energies obtained by two edge-driven methods in our simulations overlap almost completely (Fig.~\ref{fig:flexibility}b). This is due to the symmetry of the graphene homo-structures. That is different from the hetero-structure used in previous work \cite{Mandelli.prm.2018}, where push and drag (implying respectively compression and elongation) have different effects on incommensurability.

The influence of elasticity is also reflected in the effective stiffness, as suggested by Eq.~\eqref{eq3}.
In nanoscale simulations, the effect of elasticity remains  negligible due to the large in-plane stiffness of 2D materials. For a nanoscale monolayer graphene slider, in fact, the internal stiffness is $K_\mathrm{slider} \sim Y d \approx 300$~N/m ($Y$ is the Young's modulus and $d$ is the thickness of graphene, which is approximated by the interlayer distance of graphite), much larger than the external stiffness $K_\mathrm{p}$ -- typically on the order of $10$~N/m in experiments.
At the microscale, however, a thousand times larger linear size,
$K_\mathrm{slider}$ can decrease and become important. For an edge-dragged island, the internal stiffness decreases as the size  $L$ increases, $K_\mathrm{slider} \propto L^{-1}$, and the internal stiffness may become comparable to the external one \cite{Ma.prl.2015}.
In addition to the in-plane size, the thickness of the slider and the stacking will also affect the internal stiffness \cite{Roberto.nanoscale.2016} -- this is another aspect that may deserve investigation in the future.

\section{Pinning by Defects}

In the previous Sections we focused on defect-free SL islands whose interface was intact and atomically smooth. Real systems are generally more complex than this ideal case, and defects will inevitably be introduced during the synthesis and preparation of samples.
To address that kind of situation, in this chapter we discuss the influence of two common defects, vacancies and surface steps, on $U_\mathrm{eff}$ and $a_\mathrm{eff}$.

\begin{figure*}[ht!]
\centering
\includegraphics[width=0.8\linewidth]{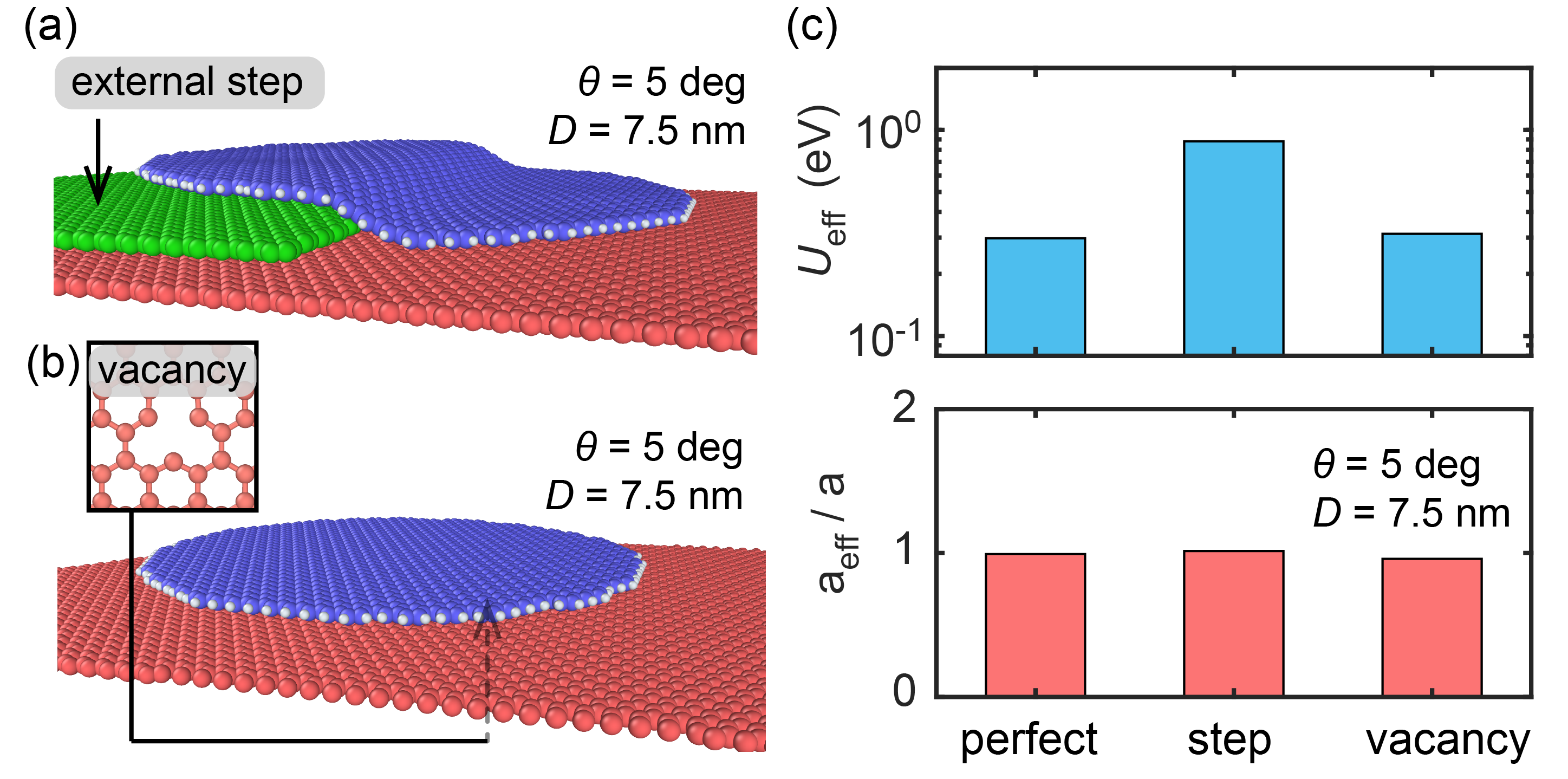}
\caption{Simulation model with (a) external step and (b) vacancy. The step is AB stacked with substrate and colored in green. The vacancy is shown in the inset.
(c) Simulation results for $U_\mathrm{eff}$ and $a_\mathrm{eff}$ for systems without defects (perfect), with one external step, and with a single vacancy. Other parameters used are $D=7.5$~nm and $\theta=5^{\circ}$.}
\label{fig:defects}
\end{figure*}

Our simulation models containing steps and vacancies are shown in Fig.~\ref{fig:defects}(a,b). One external {monolayer} step (green) is obtained by cutting the upper graphene layer of an AB stacked bilayer substrate along its armchair direction. A single vacancy (shown in inset) is introduced to the substrate within the contact region. Structures for both cases are well-optimized before the sliding simulation. 

In the results of Fig.~\ref{fig:defects}(c), we see that the difference between $a_\mathrm{eff}$ and $a$ for systems with and without step or vacancies is negligible, which again suggests using $a_\mathrm{eff} \sim a$ in general SL contacts.
On the other hand, while the energy barrier $U_\mathrm{eff}$ for the system with a single vacancy is only slightly higher than the one with perfect interface, that for the system with external step is evidently higher, an increase which will obviously enhance friction.
This is consistent with the experimental observation that the SL graphite contacts with external steps have higher friction than that of the perfect and buried step cases \cite{Wang.prl.2020}.

\section{Discussion and conclusions}

With full-atom quasi-static simulations, the dependence of effective sliding energy barrier $U_\mathrm{eff}$ and periodicity $a_\mathrm{eff}$ on size $D$, twist angle $\theta$, and island sliding direction $\Phi$ have been examined for structurally lubric graphene interfaces.
Based on these two parameters, combined with the lateral stiffness $K_\mathrm{p}$ as appropriate in a given experiment or simulation, one should be able to estimate $\eta_\mathrm{eff}$ and predict whether the sliding state will be smooth or stick-slip.

\begin{figure*}[ht!]
\centering
\includegraphics[width=0.8\linewidth]{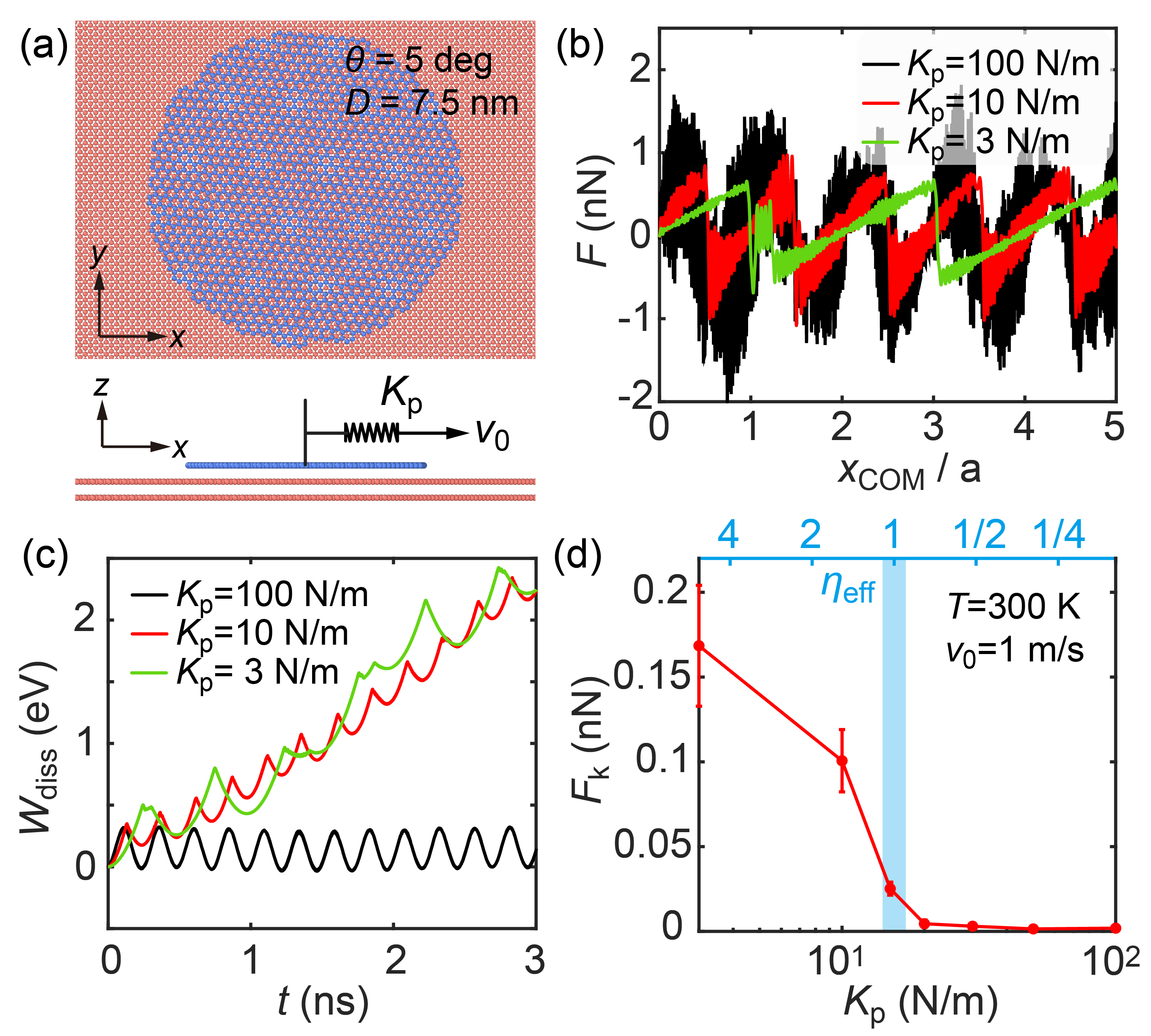}
\caption{Kinetic friction simulations.
(a) Simulation model. Parameters used are $D=7.5$~nm, $\theta=5^{\circ}$, $T=300$~K and $v_0=1$~m/s.
(b) Force traces of different $K_\mathrm{p}$ cases. Results for  $K_\mathrm{p}= 100$, $10$ and $3$~N/m correspond to smooth sliding (black), single stick-slip (red) and double-slip (green) respectively.
(c) Dissipated energy as a function of time for the three cases. Note the exceedingly small friction for $\eta_\mathrm{eff} < 1$.
(d) Kinetic friction as a function of $K_\mathrm{p}$ (lower $x$-axis) and $\eta_\mathrm{eff}$ (upper $x$-axis). The average value and error bar are estimated from three independent simulations.
The theoretically predicted transition stiffness $K_c$ is marked by the shaded region. Note the good agreement between the $\eta_\mathrm{eff}$ prediction and the actual
drop of friction.
}
\label{fig:spring}
\end{figure*}

Our prediction tool is thus ready to be tested. Of course it ought to be tested in (future) experiments. But it can also be tested right away by a direct ``realistic'' kinetic friction simulation, inclusive of temperature, sliding velocities, as well as energy dissipation. 
A room temperature simulation with $D=7.5$~nm, $\theta=5^{\circ}$ and low velocity ($v_0=1$~m/s) provides a good example.
The kinetic simulation model and set-ups (Fig.~\ref{fig:spring}a) are similar to previous work \cite{Wang.jmps.2023, Wang.rmp.2024}.
To account for energy dissipation, a Langevin thermostat is applied to the bottom layer with temperature $T=300$~K and (realistically underdamped) damping coefficient of $0.1~\mathrm{ps}^{-1}$ \cite{Wang.jmps.2023}. The time step and total simulation time used in kinetic simulations are $dt=1$~fs and $5$~ns.

Before the actual kinetic sliding simulation, in order to have a sense of the difference between the fully flexible tri-layer system (Fig.~\ref{fig:spring}a) and the fully rigid bi-layer system in Sect III, we firstly perform a quasi-static simulation. Compared to the results of the rigid model
$a_\mathrm{eff}=2.44~\mathrm{\AA}$ and $U_\mathrm{eff}=0.298$~eV in Fig.~\ref{fig:basic}(a),
here we have $U_\mathrm{eff}=0.301$~eV and $a_\mathrm{eff}=2.43~\mathrm{\AA}$ and $U_\mathrm{eff}=0.301$~eV -- the difference is negligible. In particular,
elasticity slightly lowers the barrier, as discussed in Sect VII, but the additional bottom graphene layer (required by the kinetic simulations) {compensates that.}

Shown in Fig.~\ref{fig:spring}(b-d) are the results of kinetic friction simulations with different spring stiffnesses $K_\mathrm{p}$.
As shown in Fig.~\ref{fig:spring}(b), there is a clear stick-slip when $K_\mathrm{p}=10$ and 3~N/m (corresponding to $\eta_\mathrm{eff}=1.57$ and $5.22$ respectively), as opposed to smooth sliding with $K_\mathrm{p}=100$~N/m (corresponding to $\eta_\mathrm{eff}$ = 0.16).
The simulation results very well meet the theoretical predictions -- for the underdamped low-temperature system (here $k_\mathrm{B} T/U_\mathrm{eff} \approx 0.086$),
when $\eta_\mathrm{eff}<1$, there is smooth sliding; when $1<\eta_\mathrm{eff}<4.6$, there is single stick-slip; and when $4.6<\eta_\mathrm{eff}<7.8$, there is double-slip \cite{Medyanik.prl.2006,Roth.tribolett.2010}.

The difference between smooth sliding and stick-slip further leads to significant differences in the mechanical power dissipated during sliding. We show in Fig.~\ref{fig:spring}(c) the accumulated dissipated energy $W_\mathrm{diss}$ as a function of time: for systems with $K_\mathrm{p}=10$ and 3~N/m, $W_\mathrm{diss}$ increases significantly with time, while the increase of $W_\mathrm{diss}$ for the $K_\mathrm{p}=100$~N/m system is imperceptible. This confirms that for $\eta_\mathrm{eff} < 1$ the sliding of an island is,  despite a nonzero barrier, still structurally lubric.

For completeness, we also 
{extracted for display} the kinetic friction force of the system by
$F_\mathrm{k}=\frac{\Delta W_\mathrm{diss}}{v_0 \Delta t}$ (the result has been verified to be equal to the time averaged lateral force $\langle F \rangle$).
Fig.~\ref{fig:spring}(d) shows clearly that for stick-slip cases, i.e., $\eta_\mathrm{eff}> 1$ (or $K_\mathrm{p} < 15.7$~N/m), the kinetic friction is significant; while for smooth sliding cases, the friction is much smaller. For this nanoscale simulation system, the transition stiffness dividing the two regimes is on the order of $10$~N/m, as marked by the shaded region.

Coincidentally, in many AFM-based experiments \cite{Dienwiebel.prl.2004,Liu.rsi.2007,Song.natmater.2018,Liao.natmater.2022}, the lateral stiffness of the system is also on the order of 10 N/m. This naturally
requires us to estimate $U_\mathrm{eff}$ in experiments -- typically on microscale and with large twist angles (small twist islands or flakes rotate easily back to $0^{\circ}$).
Using size scaling $U_\mathrm{eff} \sim U_i (D/a)^{1/2}$ and substituting $U_i=10^{-2}$~eV from Section III, it can be estimate that $U_\mathrm{eff}$ of a microscale system is on the order of 1 eV. Assuming $K_\mathrm{p}=10$~N/m, we can qualitatively estimate that
$\eta_\mathrm{eff} \sim 5 > 1$.
In experimental reality, the energy barrier will be generally larger due to the presence of more defects and/or contaminants. This implies that SL systems driven by AFM probes are likely to exhibit stick-slip motion unless the lateral stiffness is very much strengthened.\\

An observation that may be made here is that many SL friction experiments do not directly show stick-slip advancement of the slider, so much that superlubricity is claimed in some cases.
Strictly speaking that claim seems improper, because the measured velocity dependence, when available, is always much weaker than linear, in fact logarithmic -- and that is the hallmark of stick-slip.
The two elements, the absence of visible stick-slip and a very sublinear velocity dependence, appear contradictory at first sight. One likely explanation might be a simple lack of experimental resolution, atomic size steps being as small as they are. For very large sliders, other possibilities may involve a coexistence of many distributed pinning points, interfering with one another and transforming the advancement from stick-slip to apparently continuous.
A common feature of these seemingly contradictory cases should be a large noise. Noise is actually an observable of great importance, generally not reported and unduly neglected. The multi-pinned stick-slip should precisely differ from smooth sliding by a large increase of frictional noise. Nevertheless, the logarithmic velocity dependence of friction \cite{Wang.rmp.2024} remains a safe diagnostic and an incontrovertible proof of stick-slip in SL sliding, which we argue will experimentally occur once our criterion $\eta_\mathrm{eff} > 1$ is verified.\\

In summary, we proposed here a single PT-like parameter $\eta_\mathrm{eff}=2\pi^2 U_\mathrm{eff}/K_\mathrm{eff} a_\mathrm{eff}^2$ to describe the transition between smooth  and stick-slip  sliding of structurally lubric islands and large size interfaces.
MD simulations show systematically how the parameters vary with size, twist angle, sliding direction, lattice mismatch, elasticity, and pinning defects -- all variables that characterize real experiments.
Firstly, the sliding energy barrier $U_\mathrm{eff}$ of an island has a sublinear size scaling and is accompanied by moir\'e-sized fluctuations. For a circular island  $U_\mathrm{eff}$ decreases like $\theta^{-3/2}$ as the twist angle $\theta$ grows, and depends weakly on sliding direction. Interfacial pinning defects widely seen in experiments, especially external steps, can significantly increase the barrier.
On the other hand, the effective periodicity $a_\mathrm{eff}$ is for the cases we studied, and assuming a rigid substrate, always close to the lattice constant of the substrate $a$.
This result is attributed to the large in-plane stiffness of 2D materials leading to a relatively direction independent energy profile close to the $x=0$ minimum.
Our nanoscale simulations suggest that the island's elasticity should not be ignored, specifically at micron or larger sizes.
Elasticity reduces the intra-slider stiffness $K_\mathrm{slider}$, and that in turn reduces the  overall driving stiffness from $K_\mathrm{p}$ to $K_\mathrm{eff}$.
A smaller $K_\mathrm{eff}$ can yield $\eta_\mathrm{eff} > 1$, leading to stick-slip.

Real kinetic simulations offer a preliminary verification of the accuracy of our proposed $\eta_\mathrm{eff}$. Lastly, based on the analysis and extrapolation of simulation results, we believe that most existing SL experiments widely satisfy $\eta_\mathrm{eff}>1$.
Although stick-slip may be generally difficult to see directly in force traces, the logarithmic friction velocity dependence provides a safe diagnostic of its presence. We believe that the analysis of noise might in the future be crucial in order to further uncover the stick-slip nature of friction, when present.
On the other hand, our proposed $\eta_\mathrm{eff}$ parametrization shows that it is not impossible, even for not so 
small islands (from nano to microscales), to achieve $\eta_\mathrm{eff} < 1$ and therefore smooth sliding and negligible absolute friction (not just differential friction coefficient \cite{Wang.rmp.2024}),
despite the inevitable edge-related energy barriers.
For that goal, it will be instrumental to employ stiff drivers, sliding in directions where the slider shape has sharp wedges as opposed to flat edges.
The engineering community interested in achieving virtually frictionless smooth sliding should concentrate efforts towards reaching these conditions.

\newpage
\section*{Acknowledgments}

The author acknowledge support from ERC ULTRADISS Contract No. 834402. Support by the Italian Ministry of University and Research through PRIN UTFROM N. 20178PZCB5 is also acknowledged.
We are grateful for  discussions with A. Khosravi and A. Silva.

\bibliographystyle{unsrt}
\bibliography{ref}

\end{document}